\documentclass[sigconf]{acmart}

\usepackage{booktabs}
\usepackage{amssymb}
\usepackage{mathtools}
\usepackage{booktabs}
\usepackage{siunitx, color}
\usepackage{multirow}
\newcommand*{\functdot}{\makebox[1.5ex]{\textbf{$\cdot$}}}%

\setcopyright{rightsretained}

\copyrightyear{2017}
\acmYear{2017}
\setcopyright{acmlicensed}
\acmConference{DLRS 2017}{August 27, 2017}{Como, Italy}
\acmPrice{15.00}
\acmDOI{10.1145/3125486.3125494}
\acmISBN{978-1-4503-5353-3/17/08}

\begin{document}

\title{Music Playlist Continuation by Learning from Hand-Curated Examples and Song Features}
\subtitle{Alleviating the Cold-Start Problem for Rare and Out-of-Set Songs}

\author{Andreu Vall}
\affiliation{%
  \institution{Johannes Kepler University}
  \city{Linz}
  \country{Austria}
}
\email{andreu.vall@jku.at}
\author{Hamid Eghbal-zadeh}
\affiliation{%
  \institution{Johannes Kepler University}
  \city{Linz}
  \country{Austria}
}
\email{hamid.eghbal-zadeh@jku.at}
\author{Matthias Dorfer}
\affiliation{%
  \institution{Johannes Kepler University}
  \city{Linz}
  \country{Austria}
}
\email{matthias.dorfer@jku.at}
\author{Markus Schedl}
\affiliation{%
  \institution{Johannes Kepler University}
  \city{Linz}
  \country{Austria}
}
\email{markus.schedl@jku.at}
\author{Gerhard Widmer}
\affiliation{%
  \institution{Johannes Kepler University}
  \city{Linz}
  \country{Austria}
}
\email{gerhard.widmer@jku.at}

\renewcommand{\shortauthors}{A. Vall et al.}

\begin{abstract}
Automated music playlist generation is a specific form of music recommendation. Generally stated, the user receives a set of song suggestions defining a \emph{coherent} listening session. We hypothesize that the best way to convey such playlist coherence to new recommendations is by learning it from actual curated examples, in contrast to imposing ad hoc constraints. Collaborative filtering methods can be used to capture underlying patterns in hand-curated playlists. However, the scarcity of thoroughly curated playlists and the bias towards popular songs result in the vast majority of songs occurring in very few playlists and thus being poorly recommended. To overcome this issue, we propose an alternative model based on a song-to-playlist classifier, which learns the underlying structure from actual playlists while leveraging song features derived from audio, social tags and independent listening logs. Experiments on two datasets of hand-curated playlists show competitive performance compared to collaborative filtering when sufficient training data is available and more robust performance when recommending rare and out-of-set songs. For example, both approaches achieve a recall@100 of roughly 35\% for songs occurring in 5 or more training playists, whereas the proposed model achieves a recall@100 of roughly 15\% for songs occurring in 4 or less training playlists, compared to the 3\% achieved by collaborative filtering.
\end{abstract}

%
%




\maketitle

\section{Introduction}
\label{sec:introduction}

The ability of recommender systems to suggest coherent sets of items is particularly important in the music domain, where songs are usually grouped into listening sessions. Identifying whether a set of songs fits well together becomes then a crucial and challenging question. The study presented in \cite{cunningham_more_2006} analyzes interviews with practitioners and postings to a dedicated playlist-sharing web site. They indicate that there is a wide variety of factors (e.g., mood or purpose) that intervene in the process of compiling a playlist and suggest that the quality of a playlist and its potential coherence are not a clearly defined concept.

A common approach to playlist generation relates playlist coherence to homogeneity, meaning that the songs in a playlist should be content-wise similar (see e.g., \cite{flexer_playlist_2008,jannach_beyond_2015,knees_combining_2006,logan_content-based_2002,pohle_generating_2005}). Even though this assumption may be reasonable for some playlists, it imposes ad-hoc constraints that need not hold valid in general.

We prefer to adopt a statistical learning approach. We analyze hand-curated music playlists seeking common patterns that capture which songs fit well together. This is achieved by defining a quantitative criterion that needs to be fulfilled, on average, over playlists, thus providing a principled approach to modeling the often ambiguous concept of playlist coherence.

Collaborative Filtering (CF) methods as described in \cite{aizenberg_build_2012,bonnin_automated_2014} or the collaborative latent Markov embedding presented in \cite{chen_playlist_2012} are utilized to reveal latent patterns from hand-curated music playlists. However, CF has well-known limitations particularly detrimental for the task of playlist modeling. Most importantly, CF methods are only aware of the songs occurring in the set of training playlists. As a consequence, the songs that never occurred in the training playlists, to which we refer as ``out-of-set'' songs, can not be recommended. Also, songs that occur rarely in the training playlists are poorly represented by CF models. This problem is likely to arise due to two main reasons: firstly, the amount of carefully curated playlists is rather scarce (especially compared to the abundant--but not curated--listening logs derived from music streaming services); secondly, music consumption is biased towards popular songs \cite{celma_music_2010} and thus the vast majority of songs occur in very few playlists. Finally, CF is not aware of song characteristics, which can be informative in some cases (e.g., in order to extend playlists with a specific instrumentation or within a genre).

The aforementioned limitations of CF can be mitigated through its hybridization with content-based methods \cite{adomavicius_toward_2005}. In this line, we observe that by mining external data sources we can gather a large volume of song-level descriptions (e.g., audio, text descriptions from micro-posts or social-tagging platforms, or play counts from music streaming services). They constitute rich side-information that can be leveraged to make CF robust to rare and out-of-set songs.

In this work, we introduce a novel hybrid playlist continuation model that integrates hand-curated playlists with multi-faceted song features derived from audio, social tags and independent listening logs. Previous hybrid approaches to playlist continuation enhanced their performance through the combination of independently obtained scores by means of weighting heuristics and \mbox{re-ranking~\cite{hariri_context-aware_2012,jannach_beyond_2015}}. Instead, we blend the different sources of information into a jointly trained system, where learning is driven by the optimization of a quantitative criterion. The proposed model can then evaluate rare and even out-of-set songs, as long as song features are available, to identify songs fitting a given playlist. An implementation and data are provided for reproducibility.\footnote{\url{https://github.com/andreuvall/HybridPlaylistContinuation}}

The remainder of this paper is organized as follows. Section \ref{sec:related_work} reviews the related work. Section~\ref{sec:model} presents the hybrid playlist continuation model. The datasets of hand-curated music playlists and song features are described in Section~\ref{sec:datasets}. Section~\ref{sec:model_configurations} details the configuration of the proposed model. In Section \ref{sec:results} we elaborate on the experimental results. Finally, Section~\ref{sec:conclusion} concludes the paper.

\section{Related Work}
\label{sec:related_work}

The automated generation of coherent music playlists has often been approached from a content-based perspective. By quantifying aspects of interests in songs, a recommender system can enforce smooth transitions. For example, \cite{logan_content-based_2002} proposes to exploit timbral similarities to create a playlist of songs most similar to a given seed song. The quality of the resulting playlists is then quantified based on the criterion that the recommended songs are by the same artist, from the same album, or from the same genre as the seed. In~\cite{pohle_generating_2005}, playlist generation is treated as a ``traveling salesman problem'', where distances are defined by timbral similarities. This approach is extended in \cite{knees_combining_2006} by incorporating artist similarities computed on the basis of web-based data. The artist similarities are used to prefilter which pairwise song similarities should be calculated, resulting in an accelerated and higher-quality recommendation process. Targeting the playlist generation problem for given start and end songs, \cite{flexer_playlist_2008} propose a multi-stage approach that considers distances between all candidate songs to the start and end songs. These distances are approximated using a single Gaussian to represent the timbral features of each song.

Collaborative filtering methods have been proven powerful to mine underlying structure from, generally, user-item interactions (see e.g., \cite{ricci_recommender_2015}). In particular, CF has been applied to music playlist generation by treating playlists as users, to whom songs should be recommended. In \cite{aizenberg_build_2012}, a CF latent-factor model is tailored to mine a collection of Internet radio stations. The model features a specific latent-factor design that accounts for artist, time of the day and song adjacency. Latent variable models based on LDA \cite{blei_latent_2003} have also been applied to playlist modeling in \cite{zheleva_statistical_2010}, where a general music taste model is compared to a specific playlist model. In a similar line, \cite{chen_playlist_2012} presents the latent Markov embedding for playlist modeling. It is inspired by collaborative methods and represents songs from radio playlists into a Euclidean latent space. The latent Markov embedding puts special attention on the sequential nature of playlists and can be used to generate new playlists.

Hybrid approaches combining CF and song features have also been proposed. Playlists are modeled as random walks on song hypergraphs in \cite{mcfee_hypergraph_2012}, where the edges are derived from multimodal song features and the weights are learned from hand-curated music playlists. In \cite{hariri_context-aware_2012}, the songs in a collection of hand-curated playlists are represented by topic models extracted from song-level social tags. Frequent sequential patterns are mined at the topic level, so that given a playlist, a next topic can be predicted. To extend music playlists, the scores provided by a memory-based CF algorithm are re-ranked using the next topic predicted. Another hybrid approach for the task of playlist continuation is presented in \cite{jannach_beyond_2015}. In a first stage, suitable next songs are preselected based on the combined score of a memory-based CF algorithm, TF-IDF features derived from social tags, metadata and personal preferences. In a second stage, the song candidates are re-ranked to match the recent songs.

Similar to our approach, relating song features and collaborative patterns through neural networks has also been proposed in~\cite{van_den_oord_deep_2013}. A~convolutional neural network is trained to predict the CF factors of a song, given the log-compressed mel-spectrogram of the song. However, the two approaches are fundamentally different. Our approach integrates collaborative patterns and song features into an enhanced recommendation model. Instead, the method proposed in \cite{van_den_oord_deep_2013} emulates CF when usage data is insufficient, and its performance is naturally upper-bounded by the performance of~CF.

For a comprehensive survey on music playlist continuation, we point the interested reader to \cite{bonnin_automated_2014} or Chapter 13 in \cite{ricci_recommender_2015}.

\section{Hybrid Music Playlist Continuation}
\label{sec:model}

In this section we introduce the hybrid playlist continuation model. It is based on a song-to-playlist classifier, whose predictions can be used to recommend playlists continuations. We finally describe the evaluation methodology followed to assess its performance.

\subsection{Song-to-Playlist Classifier}
\label{sec:classifier}

Assume $T$ is a collection of hand-curated playlists where each playlist $t \in T$ is regarded as a set of songs.\footnote{As traditional CF models, the proposed model regards playlists as sets in that it does not exploit the song order. See \cite{vall_importance_2017} for a treatment where the order is considered.} Furthermore, for each playlist $t$ a continuation $t_r$ (of possibly several songs) is known and withheld as a ground-truth. Let $S$ be the set of unique songs within the collection of playlists $T$. Given a song $s \in S$, we denote its $D$-dimensional feature vector as $\mathbf{x}_s~\in~\mathbb{R}^D$. We further define a binary target vector $\mathbf{y}_s \in \{0, 1\}^{\left\vert{T}\right\vert}$ indicating the playlists to which the song belongs. 

\begin{figure}
 \centerline{
 \includegraphics[scale=0.65]{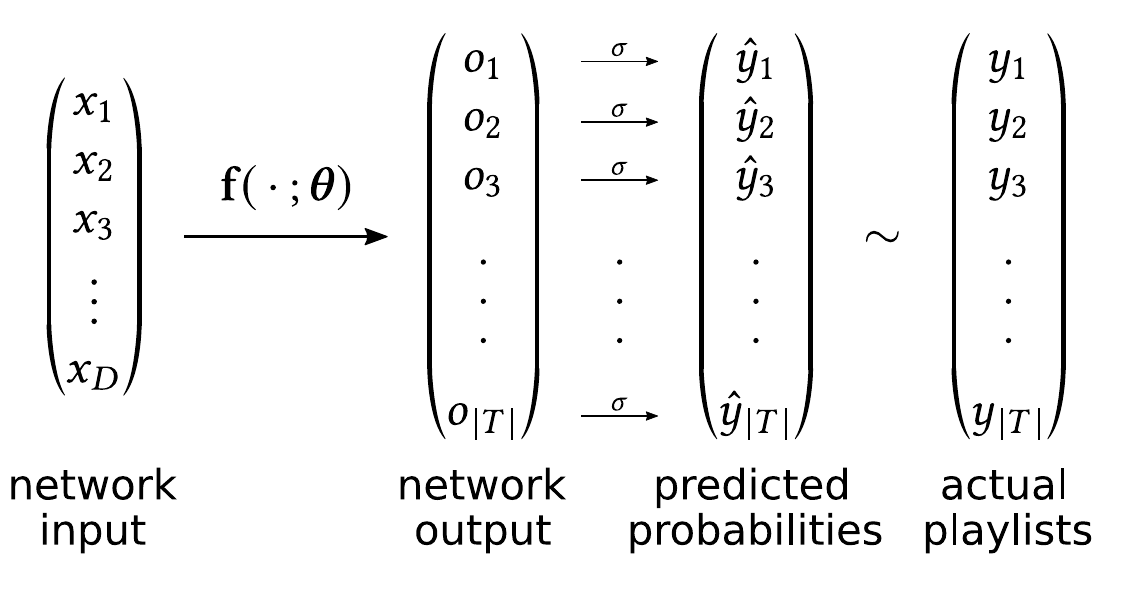}
 }
 \caption{Sketch of the neural network definition. The input is a song feature $\mathbf{x} \in \mathbb{R}^D$. The network output is pointwise passed through a logistic activation function. This yields a vector of predicted probabilities $\hat{\mathbf{y}} \in [0, 1]^{\left\vert{T}\right\vert}$ approximating the binary target $\mathbf{y} \in \{0, 1\}^{\left\vert{T}\right\vert}$, i.e., the playlists the song belongs to. The set of parameters $\boldsymbol{\theta}$ is learned from examples and its dimension depends on the network architecture.}
\label{fig:network_sketch}
\end{figure}

The proposed song-to-playlist classifier is based on a neural network $\mathbf{f}(\functdot; \boldsymbol{\theta})$, where $\boldsymbol{\theta}$ is a set of parameters common for all songs. The network takes a song feature~$\mathbf{x}_s$ as input and its output is pointwise passed through a logistic activation function yielding $\hat{\mathbf{y}}_s =  \sigma \left( \mathbf{f}(\mathbf{x}_s; \boldsymbol{\theta}) \right) \in [0, 1]^{\left\vert{T}\right\vert}$, a vector of probabilities approximating the actual binary target $\mathbf{y}_s$. Figure \ref{fig:network_sketch} sketches the neural network definition.
The song-to-playlist classifier makes as many independent decisions as playlists in the collection.\footnote{We also experimented with a softmax activation function yielding a per-song probability distribution over playlists. However, the proposed approach provided more consistent results.} Thus, the set of network parameters $\boldsymbol{\theta}$ is estimated on the training set $\{\mathbf{x}_s, \mathbf{y}_s\}_{s \in S}$ in order to minimize the following binary cross-entropy cost function
\begin{equation}
\label{eq:cost_function}
\begin{split}
\mathcal{L} \big( \boldsymbol{\theta} \mid \{\mathbf{x}_s, \mathbf{y}_s\}_{s \in S} \big) & = \\
- \sum_{s, t}
y_{s,t} & \log \big( \hat{y}_{s,t} \big) + \big( 1 - y_{s,t} \big) \log \big( 1 - \hat{y}_{s,t} \big).
\end{split}
\end{equation}
The terms $y_{s,t}$ and $\hat{y}_{s,t}$ denote the components of $\mathbf{y}_s$ and $\mathbf{\hat{y}}_s$ corresponding to playlist $t$, respectively. The dimensionality of the set of parameters $\boldsymbol{\theta}$ depends on the network architecture, which is discussed in Section \ref{sec:model_configurations}. The summation is done over all the possible song-playlist pairs, both occurring ($y_{s,t}=1$) and non-occurring ($y_{s,t}=0$) in the training playlists.\footnote{We experimented with different weighting schemes for positive and non-positive observations as suggested in \cite{hu_collaborative_2008}, but none showed a consistent improvement.}

Note that the training set $\{\mathbf{x}_s, \mathbf{y}_s\}_{s \in S}$ comprises the \mbox{song features} $\mathbf{x}_s$ and the playlist-belonging targets $\mathbf{y}_s$ (derived from hand-curated playlists). The cost function \eqref{fig:network_sketch} integrates both sources of information into a jointly learned hybrid model.

\subsection{Recommending Playlist Continuations}

Given a set of candidate songs, we use the trained song-to-playlist classifier $\mathbf{f}(\functdot; \boldsymbol{\theta^*})$ to predict the vector of probabilities $\hat{\mathbf{y}}_s \in [0, 1]^{\left\vert{T}\right\vert}$ for each candidate song $s$. The predicted vector is \emph{dense} (not \emph{sparse}), i.e., it contains the probability of song $s$ belonging to each playlist $t \in T$.\footnote{This property is also found in CF models based on latent factors (see e.g., \cite{hu_collaborative_2008}), where a dense matrix of song-playlist preferences is predicted on the basis of a sparse matrix of interactions.} 
Then, for each playlist $t \in T$, we rank all the candidate songs according to their predicted probability $\hat{y}_{s,t}$. 

The song-to-playlist classifier can evaluate any song for which a feature vector is available, even if the song does not belong to $S$. As we will see in Section~\ref{sec:results}, this enables the recommendation of out-of-set songs, and also has a positive impact on the recommendation of songs occurring only in few training playlists. 
On the other hand, as user- and factorization-based CF models, the proposed model can only extend the playlists for which it has been trained.

\subsection{Evaluation}
\label{sec:evaluation}

A large-scale on-line evaluation where users could assess the quality of the produced playlist continuations should be the preferred option (see e.g., \cite{ricci_recommender_2015}). However, this would require a complex infrastructure beyond the scope of this work. We instead opt for the standard off-line evaluation performed in \cite{aizenberg_build_2012,bonnin_automated_2014,hariri_context-aware_2012,jannach_beyond_2015}, where the ability of the model at retrieving withheld playlist continuations is assessed. Precisely, we follow \cite{aizenberg_build_2012}, where the length of each withheld continuation depends on the length of the corresponding playlist. This is contrast to \cite{bonnin_automated_2014,hariri_context-aware_2012,jannach_beyond_2015}, where only one song is withheld regardless.

We define $S^*$ by joining $S$ and the set of unique songs in the ground-truth playlist continuations. For every $s$ in $S^*$, we predict the vector of probabilities $\hat{\mathbf{y}}_s$ indicating its fit to each playlist $t \in T$. We arrange the predicted probability vectors as columns of a dense matrix of probabilities $\hat{\mathbf{Y}} \in [0, 1]^{\left\vert{T}\right\vert \times \left\vert{S^*}\right\vert}$, that has as many rows as playlists in $T$ and as many columns as songs in $S^*$.

Given a playlist $t$, we rank all the songs in $S^*$ not already included in $t$ according to the probabilities from the corresponding playlist row in $\hat{\mathbf{Y}}$. This results in a sorted list of song candidates. For each song in the withheld playlist continuation $t_r$, we compute its rank and average precision within the full list of song candidates, and we further compute its recall within the lists of top 10, top 30 and top 100 song candidates. Having the results for all the songs in all the withheld playlist continuations, we finally report the median rank, the Mean Average Precision (MAP) and the mean recall at 10, 30 and 100. Even though lists of top 30 or top 100 song candidates may seem impractical for actual applications, those and even longer lists are used to assess playlist continuations in~ \cite{aizenberg_build_2012,bonnin_automated_2014,hariri_context-aware_2012,jannach_beyond_2015}.

\section{Datasets}
\label{sec:datasets}

We use two datasets of hand-curated playlists. The ``AotM-2011'' dataset~\cite{mcfee_hypergraph_2012} is a collection of playlists derived from the Art of the Mix\footnote{\url{www.artofthemix.org}} database. Each playlist is represented by song titles and artist names, linked to the corresponding identifiers of the Million Song Dataset\footnote{\url{https://labrosa.ee.columbia.edu/millionsong}}~(MSD)~\cite{bertin-mahieux_million_2011}, where available. As we will describe in Section~\ref{sec:song_features}, the connection to the MSD is essential to our approach in order to gather additional song descriptions, from which we extract song-level features.

We also use a private playlists dataset from ``8tracks'',\footnote{\url{https://8tracks.com}} an on-line platform where users can share playlists and listen to playlists other users prepared. Similar to the AotM-2011 dataset, each playlist is represented by song titles and artist names. Mimicking the AotM-2011 dataset, we use fuzzy string matching to resolve the song titles and artist names from the 8tracks dataset against the MSD. Precisely, we adapt the code released in \cite{jansson_this_2015} for a very similar task. The matching results in roughly 2,5M song identifiers from the 8tracks dataset (many are spelling duplicates) resolved into 241,123 unique song identifiers from the MSD. The link between the 8tracks dataset and the MSD lets us gather song descriptions and makes the comparison to the AotM-2011 dataset fair.

\subsection{Playlist Continuation Sets}

The AotM-2011 dataset contains a considerable number of playlists with songs by one or very few artists. Preliminary experiments conducted on this dataset resulted in a particularly high retrieval performance when utilizing the proposed hybrid model with features derived from social tags. This result could be explained by the fact that some tags inform about the artist name, which would entail data leakage. To prevent that, we discard artist-themed playlists by keeping only the playlists with at least 7 unique artists and with a maximum of 2 songs per artist. The 8tracks dataset does not have this issue because the terms of use of the platform require that no more than 2 songs from the same artist or album may be included in a playlist. Nevertheless, we apply the same filters to both datasets for the sake of consistency. We further filter both datasets by keeping only the playlists with at least 14 songs linked to the MSD. This is to ensure that the song-to-playlist classifier has a sufficient number of songs in each playlist to learn from.

So far we have discarded complete playlists not satisfying the specified requirements. Now, for the sake of comparability, we discard the songs within playlists for which some type of feature is missing (the types of features are explained in Section \ref{sec:song_features}). As a result, the playlists are shortened and the exact amount of unique artists may be affected.

Finally, in order to set up the training and the test set, we discard all the playlists that have become shorter than 5 songs after the song filtering. We then split each playlist leaving approximately 80\% of the songs as a training example and the rest as a withheld continuation. The training examples form the training set and the withheld continuations form the test set. In this setting, a song may occur in both splits. Pure collaborative filtering approaches need to further ensure that songs occurring in the test set also occur in the training set, otherwise they can not be recommended. Our hybrid approach can deal with out-of-set songs, so this is not necessary.

The filtered AotM-2011 dataset has 2,715 playlists with 12,355 songs by 4,097 artists. The filtered 8tracks dataset has 3,272 playlists with 14,613 songs by 5,119 artists. We name the final datasets ``playlist continuation sets''. Detailed statistics regarding the distribution of unique songs per playlist, unique artists per playlist and song frequency in the datasets is reported in Table~\ref{table:stats_playlists}.
\begin{table*}
  \centering
  \captionsetup{width=0.81\textwidth}
  \caption{Descriptive statistics for the playlists within the AotM-2011 and the 8tracks playlist continuation sets. We report the distribution of the number of songs per playlist, the number of artists per playlist, and the song frequency in the dataset (i.e., the number of playlists in which each song occurs).}
  \label{table:stats_playlists}
  \begin{tabular}{rlS[table-format=1]S[table-format=1]S[table-format=1]S[table-format=2]S[table-format=3]|S[table-format=1]S[table-format=1]S[table-format=1]S[table-format=1]S[table-format=2]}
    \toprule
    && \multicolumn{5}{c|}{Training set} & \multicolumn{5}{c}{Test set} \\
    && \text{min} & \text{1q} & \text{med} & \text{3q} & \text{max} & \text{min} & \text{1q} & \text{med} & \text{3q} & \text{max} \\
    \midrule
	\multirow{3}{*}{AotM-2011}	& Songs per playlist	& 4 & 6 & 7 & 9 & 21 & 1 & 1 & 2 & 2 & 5	\\
								& Artists per playlist	& 3	& 6 & 7 & 9 & 21 & 1 & 1 & 2 & 2 & 5	\\
								& Song frequency		& 1	& 1 & 1 & 2	& 35 & 1 & 1 & 1 & 1 & 11	\\
    \midrule
	\multirow{3}{*}{8tracks}	& Songs per playlist 	& 4 & 6 & 8 & 10 & 30  & 1 & 2 & 2 & 2 & 8	\\
								& Artists per playlist	& 3	& 6 & 8 & 10 & 28  & 1 & 2 & 2 & 2 & 8	\\
								& Song frequency		& 1	& 1 & 1 & 2	 & 119 & 1 & 1 & 1 & 1 & 27	\\
    \bottomrule
  \end{tabular}
\end{table*}

\subsection{Song Features}
\label{sec:song_features}

The MSD, together with the accompanying ``Last.fm Dataset''\footnote{\url{https://labrosa.ee.columbia.edu/millionsong/lastfm}} and the ``Taste Profile Subset,''\footnote{\url{https://labrosa.ee.columbia.edu/millionsong/tasteprofile}} provide an heterogeneous collection of data for a million contemporary songs. We use song descriptions based on audio, social tags, and listening logs to extract state-of-the-art song features. For the audio content, the MSD splits songs into segments of variable length (typically under a second) and provides 12-dimensional timbral coefficients (similar to MFCCs) for each segment. Regarding the social tags, the ``Last.fm Dataset'' provides tagging activity at the song-level and at the artist-level, along with weights describing the relevance of each tag for each song and artist. Finally, the ``Taste Profile Subset'' provides user-song play counts derived from independent listening logs.\footnote{The MSD also provides high-level features such as \emph{danceability} or \emph{energy}. However, these features are not documented within the MSD nor were they within their original source, the now discontinued Echo Nest API (\url{the.echonest.com}).}

The feature extraction process is described next and is the same for the AotM-2011 and the 8tracks datasets. The extraction of song features from audio and from social tags requires the pre-estimation of models on a set of representative songs (see the details below). We prepare separate sets of songs for the AotM-2011 and for the 8tracks datasets. For each dataset, we select playlists with at least 10 songs linked to the MSD, by at least 5 artists, such that no artist has more than~2 songs in the playlist. The selected playlists are then a superset of the corresponding playlist continuation sets and we assume that the unique songs within them are representative. To prevent leaking ground-truth data we exclude the songs that appear only in the test split of the corresponding playlist continuation set. We refer to the obtained song collections as the ``development song sets''. For the AotM-2011 dataset we obtain 48,393~songs and for the 8tracks dataset we obtain 47,617~songs.

\subsubsection{Average Timbral Features}

This feature represents the average of the timbral coefficients over all the segments in a song. Each song is then described by a \mbox{12-dimensional} vector. We include it as a simple timbre-based reference.

\subsubsection{Vector-Quantized Timbral Features}

We run the $k$-means clustering algorithm on the whole pool of segment-level timbral coefficients of the development song set. We set the number of clusters to 200, thus obtaining 200 representative timbral centroids.\footnote{We extract 200-dimensional vectors for all the feature types (except for the average timbre). Our experiments indicate that 200-dimensional vectors carry rich information and fixing the dimension across features makes the comparison fair.} For each song in the playlist continuation set, we assign each frame to the closest centroid. The vector-quantized (VQ) timbral feature amounts to the count of frames the song has assigned to each centroid and therefore it is 200-dimensional. This approach has been successfully utilized in \cite{hoffman_easy_2009} for music autotagging and further investigated in \cite{seyerlehner_frame_2008} for music similarity.

\subsubsection{I-Vectors from Timbral Features}
I-vectors were first introduced in the field of speaker verification~\cite{dehak_front-end_2011}. Recently they have been successfully utilized for music similarity and music artist recognition tasks~\cite{eghbal-zadeh_i-vectors_2015,eghbal-zadeh_timbral_2015}. We build a Gaussian mixture model with 1,024 components on the entire pool of segment-level features of the development song set. Using the songs in the playlist continuation set we train the total variability space yielding \mbox{200-dimensional} i-vectors. Following the standard i-vector extraction pipeline, we further transform the i-vectors using a linear discriminant analysis model fit on the training split of the playlist continuation set.

\subsubsection{Semantic Features from Social Tags}
The embedding of words into continuous vector spaces \cite{mikolov_distributed_2013} allows us to map social tags into a semantic feature space. We gather the song-level social tags corresponding to the development song set and build a music-aware text corpus by fetching the Wikipedia\footnote{\url{https://en.wikipedia.org}} pages of the collected tags. We run the implementation of the continuous bag-of-words algorithm available in \emph{word2vec}\footnote{\url{https://code.google.com/p/word2vec}} on the text corpus to obtain a dictionary of 200-dimensional semantic features for the most relevant words. Then, for each song in the playlist continuation set we look up the words within the song-level social tags. If a tag is compound of several words (e.g., ``pop rock'') we compute the average feature value. The final song semantic feature is the weighted average of all its tags' features, where the weights are given by the relevance of each tag for the song. Likewise we compute the semantic features for the artist-level social tags, only that given a song we use the social tags related to the song's artist.

\subsubsection{Latent Factors from Listening Logs}

We factorize the user-song play counts from the ``Taste Profile Subset'' using the weighted factorization model presented in \cite{hu_collaborative_2008}, which is specifically designed for implicit feedback datasets. We use a depth of 200 factors. We discard the user latent factors, which are unrelated to our playlist continuation problem. We keep the song latent factors, that carry rich song information.

\section{Model Configurations}
\label{sec:model_configurations}

Both for the AotM-2011 and the 8tracks playlist continuation sets, we split each playlist in the training set leaving approximately 20\% of the songs for validation, which we use to determine appropriate model configurations. The ground-truth playlist continuations remain untouched until the final evaluation.

\subsection{Proposed Hybrid Model}

The proposed model is powered by the song-to-playlist classifier presented in Section \ref{sec:classifier}. Precisely, for every playlist continuation set and each type of song features we fit an independent song-to-playlist classifier, with possibly different configurations. The considered neural network architectures, hyperparameters, training strategies and feature preprocessing are detailed in Appendix~\ref{appendix:sec:proposed_hybrid_model}.

\subsection{Collaborative Filtering Baseline}
\label{sec:cf_baseline}

We compare our hybrid model to a CF baseline to assess the advantage of integrating song descriptions and hand-curated music playlists. We choose the state-of-the-art Weighted Matrix Factorization (WMF) model introduced in \cite{hu_collaborative_2008} because it is specifically designed to perform CF on implicit feedback datasets like playlists. The comparison to our model is technically simple. We just need to replace the probabilities predicted by our song-to-playlist classifier with the scores predicted by the WMF model trained on the playlist continuation sets. After that, the evaluation methodology remains valid. Further details on the WMF depth, hyperparameters and training strategy are discussed in Appendix~\ref{appendix:sec:cf_baseline}.

\section{Results}
\label{sec:results}

We evaluate the proposed playlist continuation model using the different types of song features, first as standalone features and then as combined features. Finally, we assess the performance of the proposed model to recommend rare and out-of-set songs.

Remember that the evaluation consists in, given a query playlist, retrieving the songs from its withheld continuation among all the songs in the dataset that did not occur in the query. Note that these continuations have a median length of only 2 songs (Table \ref{table:stats_playlists}) while the AotM-2011 and the 8tracks datasets have a total of 12,355 and 14,613 songs, respectively. A perfect model would rank the songs from the withheld continuations in the top positions (low ranks). An extremely poor model would rank them in the last positions (high ranks). A random model would, on average, rank them in the middle of the list of song candidates. Thus the actual rank values depend on the number of songs in each dataset.

Table \ref{table:results} reports the results of the proposed hybrid model with the different types of song features and also the results of the CF baseline, for the AotM-2011 and the 8tracks datasets. The different models have been sorted according to their median rank score.
\begin{table*}
  \centering
  \captionsetup{width=0.94\textwidth}
  \caption{Retrieval performance of the proposed model for each dataset and each type of song features, and of the CF baseline. We report the median rank, the MAP and the recall at lists of length 10, 30 and 100. The median rank compares to 12,355 song candidates for the AotM-2011 dataset and to 14,613 song candidates for the 8tracks dataset. Lower is better. For MAP and recall@\{10, 30, 100\} higher is better. As a reference, we include a random playlist continuation model where the fitness of each song-playlist pair is drawn at random from a uniform distribution in~\mbox{$[0, 1]$}.}
  \label{table:results}
  \begin{tabular}{rlS[table-format=4.1]S[table-format=0.2]S[table-format=0.2]S[table-format=0.2]S[table-format=0.2]}
    \toprule
    \textbf{dataset} & \textbf{feature} & \textbf{med rank} & \textbf{MAP} & \textbf{recall@10} & \textbf{recall@30} & \textbf{recall@100} \\
    \midrule
    AotM-2011   &   i-vectors + song tags + listening logs  	& 860       & 1.75\%    & 3.28\%    & 8.07\%    & 17.61\%   \\
                &   song tags + listening logs              	& 871       & 1.69\%    & 3.21\%    & 7.79\%    & 17.43\%   \\
                &   i-vectors + listening logs              	& 952       & 1.64\%    & 3.27\%    & 7.05\%    & 16.28\%   \\
                &   listening logs                          	& 993       & 1.64\%    & 3.10\%    & 7.37\%    & 16.32\%   \\
                &   i-vectors + song tags						& 1326      & 1.22\%    & 1.98\%    & 4.74\%    & 11.73\%   \\
                &   song tags									& 1372      & 1.25\%    & 2.34\%    & 5.26\%    & 12.99\%   \\
                &   CF											& 1444      & 1.96\%    & 3.99\%    & 7.84\%    & 14.56\%   \\
                &   artist tags                         		& 1535      & 0.68\%    & 1.14\%    & 3.11\%    & 8.68\%    \\
                &   i-vectors                           		& 2715      & 0.53\%    & 0.80\%    & 2.07\%    & 4.85\%    \\
                &   VQ timbres                          		& 3425      & 0.21\%    & 0.26\%    & 0.97\%    & 2.94\%    \\
                &   average timbres                     		& 3525      & 0.23\%    & 0.28\%    & 0.79\%    & 2.54\%    \\
                &   random                              		& 6087      & 0.11\%    & 0.20\%    & 0.28\%    & 0.79\%    \\
    \cmidrule(lr){1-7}
    8tracks     &   i-vectors + song tags + listening logs  	& 448.5     & 3.35\%    & 6.59\%    & 13.31\%   & 26.85\%   \\
                &   song tags + listening logs              	& 471       & 3.23\%    & 6.18\%    & 13.03\%   & 26.37\%   \\
                &   i-vectors + listening logs              	& 544       & 2.76\%    & 5.38\%    & 12.11\%   & 24.07\%   \\
                &   listening logs                          	& 612.5     & 2.61\%    & 4.83\%    & 10.99\%   & 23.28\%   \\
                &   i-vectors + song tags               		& 778       & 2.36\%    & 4.91\%    & 10.19\%   & 20.54\%   \\
                &   song tags                           		& 935       & 2.26\%    & 4.15\%    & 8.91\%    & 18.57\%   \\
                &   CF											& 1000      & 2.65\%    & 5.06\%    & 10.14\%   & 19.60\%   \\
                &   artist tags                         		& 1102.5    & 1.28\%    & 2.29\%    & 6.07\%    & 14.55\%   \\
                &   i-vectors                           		& 1985.5    & 0.67\%    & 1.07\%    & 2.64\%    & 7.53\%    \\
                &   VQ timbres                          		& 2897      & 0.44\%    & 0.60\%    & 1.47\%    & 4.50\%    \\
                &   mean timbres                        		& 3253.5    & 0.31\%    & 0.33\%    & 1.31\%    & 3.76\%    \\
                &   random                              		& 7320      & 0.09\%    & 0.12\%    & 0.23\%    & 0.66\%    \\
    \bottomrule
  \end{tabular}
\end{table*}

\subsection{Standalone Features}
\label{sec:standalone}

We start by analyzing the results achieved using standalone features in order to provide insights on their retrieval power for the specific task of playlist continuation. 

The performance of the different standalone features is consistent in both datasets (Table \ref{table:results}). Latent factors derived from listening logs are the most expressive feature, followed by more than 300 additional rank positions by the semantic features extracted from song-level tags. Their artist-based counterpart performs worse by approximately 200 rank positions and is the first set of song features to perform worse than the CF baseline. They are followed by far (1,200 rank positions in the AotM-2011 dataset and almost 900 rank positions in the 8tracks dataset) by the i-vectors extracted from timbral features. Other audio-based features perform worse, but clearly better than random.

The different types of features are ordered similarly in terms of their achieved MAP or their recall scores, with the exception that the CF baseline achieves a competitive recall@100 in the AotM-2011 dataset. It is also interesting to observe that the MAP values are very low for all the song features. Similar results were already observed in \cite{mcfee_natural_2011,platt_learning_2001} and could be explained by the nature of the playlist continuation problem. That is, a playlist may be continued using different songs, all of them potentially relevant, but the precision score penalizes any continuation that does not exactly match the withheld continuation. Still, we present the MAP scores for reference.

\subsection{Combined Features}
\label{sec:combined}

We extend the analysis by assessing the retrieval performance of combinations of song features. To keep the number of combinations to a moderate amount we pick only the best performing song features derived from audio (i.e., the i-vectors extracted from timbral features) and the best performing song features derived from social tags (i.e., the semantic features extracted from song-level social tags). We also use the latent factors extracted from listening logs. We evaluate all the possible combinations of pairs of song features and also the combination of the three. A combined song feature vector is simply the concatenation of the individual feature vectors. Since all the feature vectors we combine are 200-dimensional, the combinations of two features result in a 400-dimensional feature vector and the combination of the three features result in a 600-dimensional feature vector. 

Combining features improves the performance of the playlist continuation model (Table~\ref{table:results}). The most interesting case is the combination of all three types of features, which improves the median rank score by more than 100 positions compared to the best standalone feature (the latent factors extracted from listening logs). Furthermore, the recall@100 is improved by almost 1 percentage point (p.p.) in the AotM-2011 dataset and by more than 3 p.p. in the 8tracks dataset. We also observe that the gain of combining features seems to relate to their individual performance. For example, the semantic features extracted from song-level social tags perform better than the i-vectors extracted from timbral features. Then, the combination of latent factors and semantic features performs better than the combination of latent factors and i-vectors. 

The fact that the combined song features provide enhanced performance and the observation that the gain is related to their individual performance suggests that the different types of feature indeed carry different and complementary song information.

\subsection{Cold-Starting Rare and Out-of-Set Songs}

We now assess the potential advantage of the proposed hybrid model compared to the CF baseline. The proposed hybrid model outperforms the CF  baseline when it uses song features derived from song-level tags, listening logs, or any combined feature including them (Table~\ref{table:results}). The combination of all the features achieves a recall@100 almost 3 p.p. higher than CF for the AotM-2011 dataset. This difference increases to 8~p.p. in the 8tracks dataset.

\begin{figure*}[h!]
 \centerline{
 \includegraphics[scale=0.85]{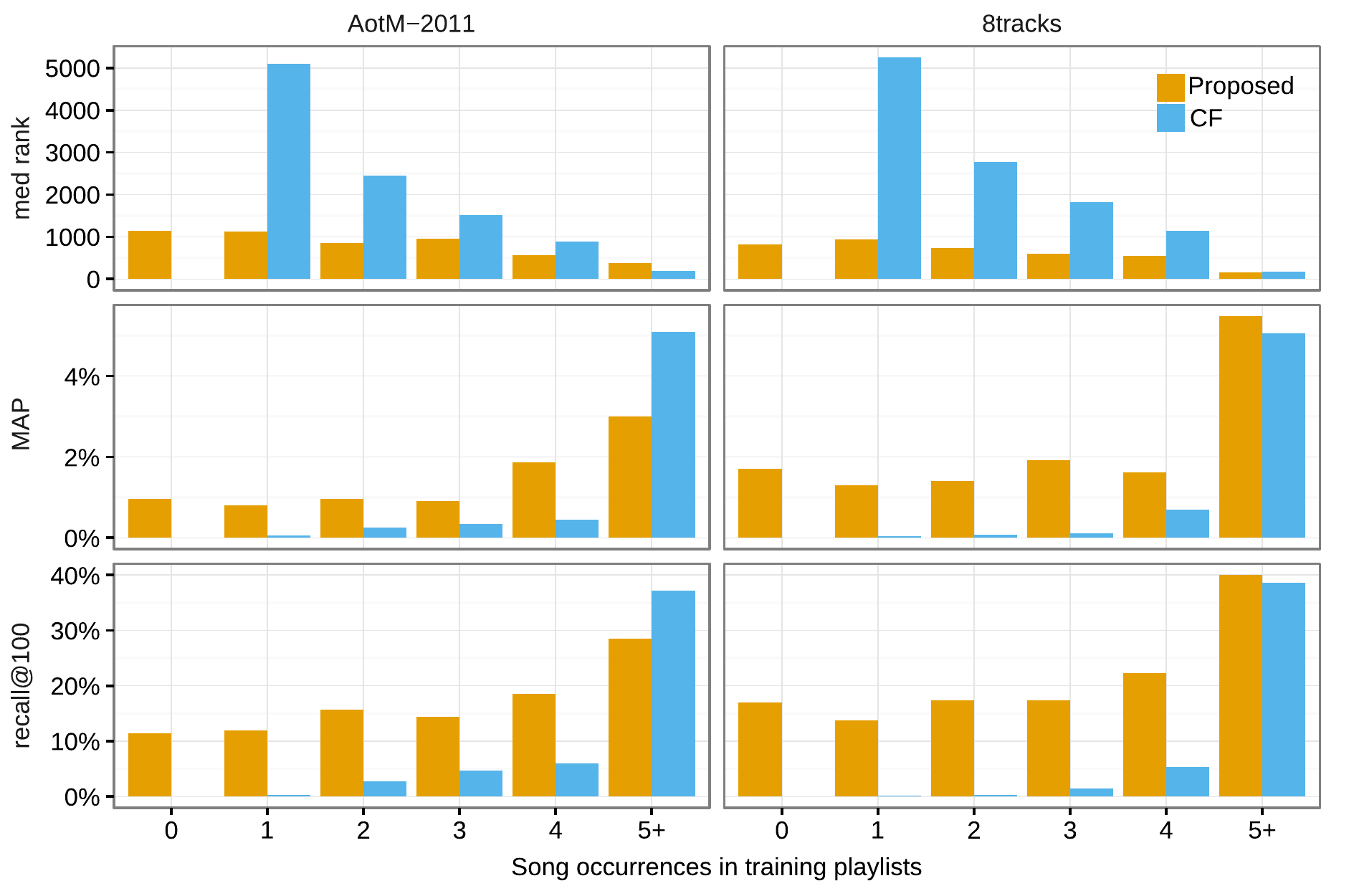}
 }
 \captionsetup{width=0.97\textwidth}
 \caption{Retrieval performance of the best performing proposed model (all features combined) and the CF baseline. For each dataset, we report the median rank, MAP and recall@100 as a function of how often the songs in the withheld continuations occurred in training playlists. For the median rank lower is better. For MAP and recall@100 higher is better.}
\label{fig:cold_start}
\end{figure*}

The performance gap between the proposed hybrid model and the CF baseline could be explained by the ability of the proposed model to deal with rare and out-of-set songs, together with the fact that in both datasets half of the songs occur only in 1 training playlist and three quarters of the songs occur only in 2 training playlists (Table \ref{table:stats_playlists}). To investigate this effect, we analyze the performance of the best performing hybrid model and the CF baseline as a function of how often the songs in the withheld playlist continuations occurred in training playlists (Figure~\ref{fig:cold_start}). The proposed hybrid model performs comparably to the CF baseline for songs with 5 or more occurrences in training playlists. For songs with 4 or less occurrences in training playists, the proposed hybrid model consistently outperforms the CF baseline in all the metrics and its performance is fairly constant regardless of the number of occurrences, even for songs that never occurred in training playists. Thus this result seems to explain the superior performance of the proposed hybrid model.

\section{Conclusion}
\label{sec:conclusion}

In this work we have introduced a hybrid music playlist continuation model that integrates collaborative information encoded in hand-curated playlists with multi-faceted state-of-the-art song-level features. In contrast to previous hybrid approaches, the proposed model fuses the different sources of information into a joint learning procedure driven by the optimization of a quantitative criterion.

We examined the performance of the model using standalone song features, as well as their combinations. Our experiments indicate that features derived from independent listening logs outperform those derived from social tags, which in turn outperform those derived from audio. Combining features improves performance further, suggesting that the different types of song features indeed carry different and complementary song information.

Most importantly, the proposed hybrid model is robust to the cold-start problem for rare and even out-of-set songs. Indeed, our experimental results confirm that the proposed model consistently outperforms the CF baseline when data is scarce. If data is abundant, the proposed model performs comparably to the CF baseline.

\newpage
\bibliographystyle{ACM-Reference-Format}
\bibliography{vall_etal_hybrid_playlist_continuation.bib}

\appendix

\section{Model Configurations}
\label{appendix:sec:model_configurations}

\subsection{Proposed Hybrid Model}
\label{appendix:sec:proposed_hybrid_model}

We conducted an initial exploration of architectures by evaluating networks with $\{2, 3, 4\}$ hidden layers, $\{50, 100, 200, 500\}$ hidden units, learning rate values in $\{0.1, 0.5, 1.0\}$ and batch sizes of $\{10, 50, 100, 200\}$ songs. We also experimented with the hyperbolic tangent, the logistic function and the rectifier as activation functions for the hidden layers. We did not perform all the combinations of the aforementioned parameter values, but used the preliminary results to narrow down the actual parameter search space.

Given the results of the preliminary analysis, we systematically explored all the combinations of networks with $\{2, 3\}$ hidden layers and with $\{50, 100, 200\}$ hidden units. We fixed the learning rate to 0.5, the batch size to 50 songs and the hyperbolic tangent as the activation function for hidden layers. Recall that the output layer of the network is passed through logistic functions~\mbox{(see Section \ref{sec:classifier}).}

We use batch normalization \cite{ioffe_batch_2015}. We also experimented with different dropout probabilities~\cite{srivastava_dropout:_2014} and with $L1$ and $L2$ regularization to prevent overfitting. We finally decided to use dropout with probabilities 0.1 and 0.5 at the input layer and the hidden layers, respectively.

The features were preprocessed. Namely, the average timbral features, the i-vectors from timbral features, the semantic features from social tags, the latent factors from listening logs and the combined features were standardized and $L2$-normalized. The vector-quantized timbral features were only $L1$-normalized according to their histogram-like nature.

The networks were optimized to minimize the cost function~\eqref{eq:cost_function} using AdaGrad\footnote{Since AdaGrad re-scales the learning rate at every update, setting the learning rate to 0.5 actually refers to setting its initial value.} \cite{duchi_adaptive_2011} with Nesterov momentum~\cite{nesterov_method_1983}. We trained for a maximum of 1,000 epochs but stopped before if the cost function was not significantly minimized during 100 epochs. The cost function drove the optimizer, but the best model was chosen on the basis of the highest recall achieved on the validation set. We also used the recall on the validation set to decide an appropriate number of epochs for the final training on the entire training set. Table~\ref{table:settings} reports the final configuration of each network. We implemented the networks using Lasagne~\cite{dieleman_lasagne:_2015}, which is built on top of Theano~\cite{theano_development_team_theano_2016}.
\begin{table}[h!]
  \centering
  \caption{Final network configuration for each playlist continuation set and each set of features. The learning rate is set to 0.5, the batch size to 50 songs and the hyperbolic tangent is the activation function for all the hidden layers. We use batch normalization and dropout with probabilities 0.1 and 0.5 in the input layer and the hidden layers, respectively.}
  \small
  \label{table:settings}
  \begin{tabular}{rlS[table-format=1]S[table-format=3]S[table-format=3]}
    \toprule
    \textbf{dataset} & \textbf{feature} & \textbf{layers} & \textbf{units} & \textbf{epochs}  \\
    \midrule
    AotM-2011   &   average timbres                         & 2 & 50    & 100   \\
                &   VQ timbres                              & 3 & 50    & 300   \\
                &   i-vectors                               & 3 & 50    & 200   \\
    \cmidrule[0.5pt](lr){2-5}
                &   song tags                               & 3 & 100   & 250   \\
                &   artist tags                             & 3 & 50    & 600   \\
    \cmidrule[0.5pt](lr){2-5}
                &   listening logs                          & 3 & 100   & 310   \\
    \cmidrule[0.5pt](lr){2-5}
                &   i-vectors + song tags                   & 3 & 50    & 500   \\
                &   i-vectors + logs              & 3 & 100   & 230   \\
                &   song tags + logs              & 3 & 100   & 200   \\
                &   i-vectors + song tags + logs  & 3 & 100   & 150   \\
    \cmidrule(lr){1-5}
    8tracks     &   average timbres                         & 3 & 50    & 200   \\
                &   VQ timbres                              & 3 & 50    & 450   \\
                &   i-vectors                               & 3 & 50    & 500   \\
    \cmidrule[0.5pt](lr){2-5}
                &   song tags                               & 2 & 100   & 408   \\
                &   artist tags                             & 2 & 100   & 500   \\
    \cmidrule[0.5pt](lr){2-5}
                &   listening logs                          & 2 & 50    & 540   \\
    \cmidrule[0.5pt](lr){2-5}
                &   i-vectors + song tags                   & 3 & 100   & 300   \\
                &   i-vectors + logs              & 3 & 100   & 360   \\
                &   song tags + logs              & 3 & 100   & 520   \\
                &   i-vectors + song tags + logs  & 3 & 100   & 360   \\
    \bottomrule
  \end{tabular}
\end{table}

\subsection{Collaborative Filtering Baseline}
\label{appendix:sec:cf_baseline}

The Weighted Matrix Factorization (WMF) model introduced in~\cite{hu_collaborative_2008} generally mines user-item interactions while leveraging the intensity of the interactions (e.g., the number of clicks on websites, or the play counts on on-line streaming services). Modeling playlists is a slightly different problem because the examples consist of binary values without intensity information. Thus, the default weighting scheme proposed in \cite{hu_collaborative_2008} is not suited for our task. We used the validation sets to experiment with different weights for the observed playlist-song interactions and found that assigning them a weight of 2 yielded best results. We also experimented with different weights for the $L2$-regularization term and decided to use a factor of 10. We use the implementation provided in \cite{frederickson_fast_2017}.

\end{document}